\documentclass[aps,pre,amsmath]{revtex4}
\usepackage{amsmath}
\usepackage[cp1251,koi8-r]{inputenc}
\usepackage[english]{babel}
\usepackage{epsfig}
\usepackage{graphics}

\newcommand{\be}{\begin{equation}}
\newcommand{\ee}{\end{equation}}
\newcommand{\bn}{\begin{eqnarray}}
\newcommand{\en}{\end{eqnarray}}
\newcommand{\bd}{\begin{displaymath}}
\newcommand{\ed}{\end{displaymath}}
\newcommand{\bnn}{\begin{eqnarray*}}
\newcommand{\enn}{\end{eqnarray*}}
\def\Ref#1{(\ref{#1})}

\begin{document}
\inputencoding{cp1251}

\title{Universal Approach to Overcoming Nonstationarity, Unsteadiness and Non-Markovity
of Stochastic Processes in Complex Systems }

\author{Renat M. Yulmetyev${}^{1,}$\thanks{e-mail:rmy@dtp.ksu.ras.ru},
Anatolii V. Mokshin${}^{1,}$\thanks{e-mail:mav@dtp.ksu.ras.ru},
and Peter H\"anggi${}^{2}$ }
\address{$^{1}$Department of Physics, Kazan State
Pedagogical University, Kazan, Mezhlauk 1, 420021
Russia \\
$^{2}$Department of Physics, University of Augsburg,
Universit\"atsstrasse 1, D-86135 Augsburg, Germany}

\begin{abstract}
In present paper we  suggest a new universal approach to study
complex systems by microscopic, mesoscopic and macroscopic
methods. We  discuss new possibilities of extracting information
on  nonstationarity, unsteadiness and non-Markovity of discrete
stochastic processes in complex systems. We consider statistical
properties of the fast, intermediate and slow components of the
investigated processes  in complex systems within the framework of
microscopic, mesoscopic and macroscopic approaches separately.
Among them  theoretical analysis is carried  out by means of local
noisy time-dependent parameters and the conception of a
quasi-Brownian particle (QBP) (mesoscopic approach) as well as the
use of wavelet transformation of the initial row time series. As a
concrete example we examine the seismic time series data for
strong and weak earthquakes in Turkey ($1998,1999$) in detail, as
well as technogenic explosions. We propose a new way of possible
solution to the problem of forecasting strong earthquakes
forecasting. Besides we have found out that an unexpected
restoration of the first two local noisy parameters in weak
earthquakes and technogenic explosions is determined by
exponential law. In this paper we have also carried out the
comparison and have discussed the received time dependence of the
local parameters for various seismic phenomena.

\end{abstract}

\maketitle

%%%%%%%%%%%%%%%%%%%%%%%%%%%%%%%%%%%%%
\section{Introduction}\label{Sec_1}
%%%%%%%%%%%%%%%%%%%%%%%%%%%%%%%%%%%%%
Nonstationarity, unsteadiness and non-Markovity are the most
common essential peculiarities of stochastic processes in nature.
The existence of the similar properties  creates significant
difficulties for the theoretical analysis of real complex  systems
\cite{Stanley}. At present, methods, connected with localization
of registered or calculated parameters for the quantitative
account of the dramatic changes caused by the fast alternation of
the behavior modes and intermittency, came into use. For example,
the time behavior of the local (scale) Hurst exponents was defined
in the recent work of Stanley \textit{et al.} to study
multifractal cascades in heartbeat dynamics \cite{Stanley} and to
analyze and forecast earthquakes and technogenic explosions in
Ref. \cite{Yulmetyev1}. The application  of the local
characteristics allows one to avoid difficulties connected with
nonergodicity of the investigated system and gives a possibility
to extract additional valuable information on the hidden
properties of real complex systems. From the physical point of
view, this approach resembles the use of nonlinear equations of
generalized hydrodynamics with the local time behavior of
hydrodynamical and thermodynamical parameters and characteristics.

It is well known that one of the major problems of seismology is
to predict the beginning of the main shock. Though science seems
still to be far from the guaranteed decision of this problem there
exist some interesting approaches based on the peculiar properties
of precursory phenomena
\cite{Sornette1,Igarashi,Sornette2,Sornette3,Bufe,Bak,Sornette4}.
Another important problem is recognition and differentiation of
weak earthquakes and technogenic underground explosions signals.
One of useful means of solving this problem is the defining of
corresponding local characteristics \cite{Stanley,Yulmetyev1}.

In the present work we  suggest a new universal description of
real complex systems by means of the microscopic, mesoscopic and
macroscopic methods. We start with a macroscopic approach based on
the kinetic theory of discrete stochastic processes and the
hierarchy of the chain of finite-difference kinetic equations for
the discrete time correlation function (TCF) and memory functions
\cite{Yulmetyev1,Yulmetyev2,Yulmetyev3}.

The mesoscopic phenomena of the so-called "soft matter"\ physics,
embracing a diverse range of system including liquid crystals,
colloids, and biomembranes, generally involve some form of
coupling of different characteristic time- and length-scales.
Computational modelling of such multi-scale effects requires a new
methodology applicable beyond the realm of traditional techniques
such as \textit{ab initio} and classical molecular dynamics (the
methods of choice in the microscopic regime), as well as phase
field modelling or the lattice-Boltzmann method (usually concerned
with the macroscopic regime). As for complex systems, we propose
to consider intermediate and slow processes within a unified
framework of mesoscopic approach: by means of local time behavior
of the local relaxation and kinetic parameters, local
non-Markovity parameters and so on. For this purpose we introduce
the notion of quasi-Brownian motion in a complex system by
coarse-grained averaging of the initial time series on the basis
of wavelet transformation.

As an example we consider here the local properties of relaxation
or noise parameters for the analysis of seismic phenomena such as
earthquakes and technogenic explosions. The layout of the paper is
as follows. In Sec. \ref{Sec_Loc} we describe in brief the
stochastic dynamics of time correlation in complex systems by
means of discrete non-Markov kinetic equations. Basic equations
used for these calculations are presented here. The local noise
parameters are defined in Sec. \ref{local_parameters}. Section
\ref{definition of local parameters} contains results obtained by
the local noise parameter procedure for the case of seismic
signals. The models of the time dependence of the local parameters
are given in Sec. \ref{model}. The basic conclusions are discussed
in Sec. \ref{conclusion}.

%%%%%%%%%%%%%%%%%%%%%%%%%%%%%%%%%%%%%%%%%%%%%%%%%%%%%%%%%%%%%%%%%%%%%
\section{The basic definitions in kinetic description of discrete
stochastic processes}\label{Sec_Loc}
%%%%%%%%%%%%%%%%%%%%%%%%%%%%%%%%%%%%%%%%%%%%%%%%%%%%%%%%%%%%%%%%%%%%%
\subsection{MACROSCOPIC DESCRIPTION IN THE  ANALYSIS OF STOCHASTIC
PROCESSES}\label{sub_1}
%%%%%%%%%%%%%%%%%%%%%%%%%%%%%%%%%%%%%%%%%%%%%%%%%%%%%%%%%%%%%%%%%%%%%
 A lot of different existent processes, such as economical,
metheorogical, gravimetrical and other, are registered as discrete
random series $x_{i}$ of some variable $X$. This random variable
$X$ can be written as an array of its values \be X=\{x(T),
\enspace x(T+\tau),\enspace x(T+2\tau),\enspace...,\enspace
x(T+k\tau),\enspace...,\enspace x(T+(N-1)\tau)\}. \label{series}
\ee Here a time step (or a time interval) $\tau$ is a constant,
$T$ is the time when the registration of the signal begins,
$(N-1)\tau$ is the duration of the signal detection.

The average value $\langle x \rangle$ and fluctuations $\delta
x_{j}$ are defined by the following expressions, correspondingly
\be \langle x \rangle = \frac{1}{N} \sum_{j=0}^{N-1} x(T+j\tau), \
\ \delta x_{j}=\delta x(T+j\tau)=x(T+j\tau)-\langle x \rangle.
\label{fluct} \ee

From the fluctuations of the considered random variable $\delta
x_{j}$ we can form $k$-component state vector of the system's
correlation state \be \textbf{A}_{k}^{0}=\textbf{A}_{k}^{0}(0) =
(\delta x(T), \enspace \delta  x(T+\tau), \enspace ..., \enspace
\delta x(T+(k-1)\tau)) =(\delta x_{0}, \enspace \delta x_{1},
\enspace ..., \enspace \delta x_{k-1}). \label{initial_vector} \ee

The time dependence of the correlation state vector \textbf{A} can
be represented as a discrete  $m$-step shift \be
\textbf{A}_{m+k}^{m}=\textbf{A}_{m+k}^{m}(t)=(\delta x(T+m\tau),
\enspace \delta  x(T+(m+1)\tau), \enspace ..., \enspace \delta
x(T+(m+k+1)\tau)) =(\delta x_{0}, \enspace \delta x_{1}, \enspace
..., \enspace \delta x_{m+k-1}). \label{intermediate_vector} \ee
Then by analogy with the papers
\cite{Yulmetyev1,Yulmetyev2,Yulmetyev3} we can write the following
normalized TCF \be M_{0}(t)=\frac{\langle \textbf{A}_{N-1-m}^{0}
\textbf{A}_{N-1}^{m} \rangle}{\langle \textbf{A}_{N-1-m}^{0}
\textbf{A}_{N-1-m}^{0} \rangle} = \frac{\langle
\textbf{A}_{N-1-m}^{0}(0) \textbf{A}_{N-1}^{m}(t) \rangle}{|
\textbf{A}_{N-1-m}^{0}(0)|^{2} } = \frac{\langle
\textbf{A}_{N-1-m}^{0}(0) U(t=T+m\tau,T) \textbf{A}_{N-1-m}^{0}(0)
\rangle}{| \textbf{A}_{N-1-m}^{0}(0)|^{2}}, \label{initial_TCF}
\ee where angular brackets indicate the scalar product of the two
state vectors. On the other hand, the time dependence of the
vector ${\textbf{A}}_{N-1-m}^{m}(T+t)$, $t=m\tau$, can be
represented formally with the help of the evolution operator
$U(t',t)$ as follows: \be \textbf{A}_{N-1}^{m}(T+t)=
U(T+m\tau,T)\textbf{A}_{N-1-m}^{0}(T)=U(t,0)\textbf{A}_{N-1-m}^{0}(0).
\ee The last one has the property: $U(t,t)=1$. Actually, one can
write down formal discrete equation of motion with the use of the
evolution operator $U(t',t)$ (see Appendix A for more details).

It was shown in Refs. \cite{Yulmetyev1,Yulmetyev2,Yulmetyev3} that
the finite-difference kinetic equation of a non-Markov type for
TCF $M_{0}(t)$ can be written by means of the technique of
projection operators of Zwanzig'-Mori's  type \cite{Zwanzig,Mori}
as \be \frac{\triangle M_{0}(t)}{\triangle t} = \lambda_{1}
M_{0}(t)-\tau \Lambda_{1} \sum_{j=0}^{m-1}
M_{1}(j\tau)M_{0}(t-j\tau). \label{first_eq} \ee Here the first
order memory function $M_{1}(j\tau)$ appears, $\lambda_{1}$ is the
eigenvalue of Liouville's quasi-operator $\hat{\mathcal{L}}$ and
$\Lambda_{1}$ is the relaxation noise parameter, which are
characteristics of the investigated process. Possible methods of
defining quasioperator $\hat{\mathcal{L}}$ are presented in
Appendix A [see Eqs. \Ref{Liuv_1},\Ref{Liuv_2} and \Ref{Liuv_3}].
It should be recorded that Eq. \Ref{first_eq} is the first kinetic
finite-difference equation for initial TCF $M_{0}(t)$. With the
use of the same procedure of projection operator we can obtain the
chain of kinetic finite-difference equations of the following
form: \be \frac{\triangle M_{i-1}(t)}{\triangle t} = \lambda_{i}
M_{i-1}(t)-\tau \Lambda_{i} \sum_{j=0}^{m-1}
M_{i}(j\tau)M_{i-1}(t-j\tau), \ \ i=1,2,3,... . \label{chain} \ee
Here $M_{i}(j\tau)$ is the memory function of the $i$th order,
whereas $\lambda_{i}$ and $\Lambda_{i}$ are noise parameters: \be
\lambda_{n}=i\frac{\langle
\textbf{W}_{n-1}\hat{\mathcal{L}}\textbf{W}_{n-1}\rangle}{|\textbf{W}_{n-1}|^{2}};
\ \ \Lambda_{n}=i\frac{\langle
\textbf{W}_{n-1}\hat{\mathcal{L}}\textbf{W}_{n}\rangle}{|\textbf{W}_{n-1}|^{2}}.
\label{noise_parameters} \ee Here  $\textbf{W}_{n}$ are the
dynamical orthogonal variables, obtained by the Gram-Schmidt
orthogonalization procedure \be
\langle\textbf{W}_{n},\textbf{W}_{m}\rangle = \delta_{n,m}\langle
|\textbf{W}_{n}|^{2} \rangle, \nonumber \label{Gram_Schmidt} \ee
where $\delta_{n,m}$ is the Kronecker's symbol, \be
\textbf{W}_{0}=\textbf{A}_{k}^{0}(0), \ \
\textbf{W}_{1}=[i\hat{\mathcal{L}}-\lambda_{1}]\textbf{W}_{0}, \nonumber\\
\ee
\be
\textbf{W}_{2}=[i\hat{\mathcal{L}}-\lambda_{2}]\textbf{W}_{1}-\Lambda_{1}\textbf{W}_{0},
... . \label{Dynamic_var}
\ee

From Eq. \Ref{Dynamic_var} it is obvious that in the cited
Gram-Schmidt procedure from each new vector of state one should
subtract the projection on all the previous vectors. Thereafter
the orthogonalization \Ref{Gram_Schmidt} is complete.

A chain of integro-differential equations \Ref{chain} arise as a
result of the use of projection operator technique to define
different correlation functions in physical problems
\cite{Zwanzig,Mori} (for example, TCF of density fluctuation in
Inelastic Neutron Scattering \cite{Yulmetyev_Mokshin,Scopigno} and
Light Scattering investigations, velocity autocorrelation function
and others can be received in the specified way). However, the
quest for the physically based way of closing the chain of
equations and finding the spectra of the initial TCF are essential
moments in these challenges. Here the situation is different.
Namely, the initial TCF can be calculated directly from the
experimental data. Then memory functions $M_{i}(t)$ and noise
parameters $\lambda_{i}$, $\Lambda_{i}$ are similarly calculated
from the experiment. All these functions and parameters make it
possible to carry out the detailed analysis of the random process.
The noise parameters $\lambda_{i}$ and $\Lambda_{i}$ are the
relaxation characteristics of the experimental time series, which
contain information of various modes passing and changing. The
macroscopic approach presented above is based on the calculation
of memory functions, power spectra, dynamic variables, relaxation
parameters. It suggests the investigation of the system as a
single whole. The global characteristics calculated on the basis
of all the time series contain hidden information about various
modes of the system behavior. As a rule, this information is
difficult to extract and analyze. For this reason it is necessary
to develop a mesoscopic description and introduce a local noise
and relaxation parameters $\lambda_{i}$ and $\Lambda_{i}$.

%%%%%%%%%%%%%%%%%%%%%%%%%%%%%%%%%%%%%%%%%%%%%%%
\subsection{MESOSCOPIC DESCRIPTION IN ANALYSIS OF STOCHASTIC PROCESSES}
%%%%%%%%%%%%%%%%%%%%%%%%%%%%%%%%%%%%%%%%%%%%%%%
It is well known, the mesoscopic conception is one of the possible
ways to deal with random processes in complex systems. It consists
of extension  of the domain of the dynamical equations mesoscopic
variables and of introduction of a some local time interval. Such
quantity as $X$ (it can be particle position, mass density and so
on) is defined on the mesoscopic space \cite{mesoscop}. In
addition we introduce number $M$ as its extensive quantity. This
number $M$ should satisfy the condition: \be 1 \ll M \ll N,
\label{condition} \ee where $N$ is the length of the initial
experimental sampling. Let us take a working window of the fixed
length $M$. By superposing this window on the initial sampling
$X$, we choose all elements, incoming into it, as a separate
sampling $\bar{\xi}_{0}$. Further, let us execute one time step
$\tau$ shift of this working window to the right and obtain
another local sampling of the length $M$. Executing this procedure
($N-M+1$) times, one can obtain the same quantity of the local
samplings of length $M$: \bn
\bar{\xi}_{0}&=&\bar{\xi}_{0}\{x(T),x(T+\tau),x(T+2\tau), \ldots,
x(T+(M-1)\tau)\},  \nonumber \\
\bar{\xi}_{1}&=&\bar{\xi}_{1}\{x(T+\tau),x(T+2\tau),x(T+3\tau),
\ldots,
x(T+M\tau)\},  \nonumber\\
& &...,  \nonumber\\
\bar{\xi}_{N-M}&=&\bar{\xi}_{N-M}\{ x(T+(N-M-1)\tau),
x(T+(N-M)\tau),x(T+(N-M+1)\tau) \ldots, x(T+(N-1)\tau) \}.
\label{loc_last}\en The obtained local samplings $\bar{\xi}_{i}$
form the array, which represents the time dynamics of the
investigated process, \be \bar{\xi}(t')= \{\bar{\xi}_{0},
\bar{\xi}_{1}, \bar{\xi}_{2}, \ldots,\bar{\xi}_{i}, \ldots,
\bar{\xi}_{N-M} \}. \label{new_series} \ee Then, in accordance
with the procedure described in the last subsection \Ref{sub_1},
we can define fluctuations by Eq. \Ref{fluct}, calculate
 TCF with the help of Eq. \Ref{initial_TCF}, calculate memory
functions and parameters $\lambda_{i}$ and $\Lambda_{i}$ by Eqs.
\Ref{noise_parameters} for every sampling $\bar{\xi}_{i}$.
However, parameters $\lambda_{i}$ and $\Lambda_{i}$ will be
characteristics of concrete $j$th sampling only. To characterize
the local properties of the initial time data $X$, it is
convenient to represent their local time dependence in the
following way: \bn \lambda_{i}(t')=\{\lambda_{i}(T+(M-1)\tau),
\lambda_{i}(T+M\tau), \lambda_{i}(T+(M+1)\tau), \ldots,
\lambda_{i}(T+(N-M)\tau)
\},\nonumber\\
\Lambda_{i}(t')=\{\Lambda_{i}(T+(M-1)\tau), \Lambda_{i}(T+M\tau),
\Lambda_{i}(T+(M+1)\tau), \ldots, \Lambda_{i}(T+(N-M)\tau) \}.
\en

Operating in the similar way, one can execute cross-over from the
macroscopic description of the whole system to the mesoscopic one.
The offered approach is very convenient for the description and
analysis of non-stationary stochastic processes. It allows to
depart from the global macro-characteristics, which carry only
averaged minor information about the whole investigated process,
and to turn to the stochastic description with the use of the
local characteristics, and, as a result, to execute a more
detailed analysis of various dynamic states of the system.

%%%%%%%%%%%%%%%%%%%%%%%%%%%%%%%%%%%%%%%%%%%%%%%%%%%%%%%%%%%%
\subsection{CONCEPTION OF ONE-DIMENSIONAL DYNAMICS OF QUASI-BROWNIAN PARTICLE}

Let us consider the motion of a large Brownian particle in a dense
medium composed of light molecules and restrict it by a simple
one-dimensional case. The coordinate $x_{i}$ and the velocity
$v_{i}$ are random variables of a Brownian particle. The quantity
$\tau$ represents average time between the two successive
collisions of liquid molecules, $T$ is the initial moment of time.
the variable $M$ characterizes a local size (mass) of a Brownian
particle. It is obvious, that a Brownian particle must have a
larger mass in comparison with liquid particles ($M\gg1$),
therefore it is more inert. Then it is convenient to define the
coordinate of a Brownian particle at moment $t'$ as an average
value of sampling $\bar{\xi}_{i}$. For example, we obtain from
$\bar{\xi}_{0}$: \be y_{0}=\frac{x(T)+x(T+\tau)+x(T+2\tau)+\ldots+
x(T+(M-1)\tau)}{M} =\frac{1}{M}\sum_{j=0}^{M-1}x(T+j\tau).
\label{average_1} \ee The quantity $y_{0}$ defines the coordinate
of "the center of mass" of a Brownian particle at the initial time
moment $t'=0$. By analogy, it can define the position of a
Brownian particle at the next time moment $t'=\tau$ and so on. As
a result, we obtain a new time discrete series $Y(t')$ as: \be
Y(t')=\{y_{0},y_{1},y_{2},\ldots,y_{N-M} \}
\label{position_array}. \ee The velocity of a Brownian particle is
\be v_{i}=\frac{y_{i+1}-y_{i}}{\tau}. \label{velocity} \ee So, for
example, for the initial velocity $v_{0}$ we obtain from Eqs.
\Ref{loc_last}, \Ref{average_1} and \Ref{velocity}:  \be
v_{0}=\frac{y_{1}-y_{0}}{\tau}=\frac{x(T+M\tau)-x(T)}{M\tau}. \ee
Obviously, \textit{if the larger one is $M$, the smaller one is
the velocity of a Brownian particle}. So, we obtain a discrete set
of velocity values for a Brownian particle at equally small time
interval $\tau$ \be V(t')=\{v_{0}, v_{1}, v_{2}, \ldots, v_{N-M-1}
\}. \label{velocity_array} \ee

Eqs. \Ref{position_array} and \Ref{velocity_array} define time
dependence of random variables $Y(t')$ and $V(t')$.

%%%%%%%%%%%%%%%%%%%%%%%%%%%%%%%%%%%%%%%%%%%%%%%%%%%%%%%%%%%%%%%%

\subsubsection{Generalized Langevin equation with discrete time}

%%%%%%%%%%%%%%%%%%%%%%%%%%%%%%%%%%%%%%%%%%%%%%%%%%%%%%%%%%%%%%%%

In accordance with  Eq. \Ref{fluct} we define the correlation
state vector, components of which are the fluctuations of the
particle position, \be \textbf{B}_{k}^{0}=\{\delta y_{0},\ \delta
y_{1},\ \delta y_{3}, \ldots,\ \delta y_{k-1} \}.
\label{first_vec} \ee Then, by analogy to Eq.
\Ref{intermediate_vector} time dependence of the vector \textbf{B}
can be considered a discrete $m$-step time shift. For the vector
\textbf{B} the following normalized TCF can be written with the
help of Eq. \Ref{initial_TCF} \be b(t)= \frac{\langle
\textbf{B}_{N-1-m}^{0} \textbf{B}_{N-1}^{m} \rangle}{|
\textbf{B}_{N-1-m}^{0}|^{2} }. \ee The last one describes the time
correlation of the two different correlation states of the system.

Now we introduce the linear projection operator in Euclidean space
of the state vectors \be Q
\textbf{B}=\frac{\textbf{B}(0)\rangle\langle\textbf{B}(0)
\textbf{B}(t)\rangle}{|\textbf{B}(0)|^{2}}=\textbf{B}(0)\rangle\
b(t) ,\ Q=\frac{\textbf{B}(0)\rangle \langle\textbf{B}(0)}{\langle
\textbf{B}(0) \textbf{B}(0)\rangle}. \label{oper} \ee This
operator has the necessary property of idempotency $Q^{2}=Q$. The
existence of projection operator $Q$ allows to introduce the
mutually-supplementary projection operator $R$ as follows: \be
R=1-Q,\ R^{2}=R,\ QR=RQ=0. \ee It is necessary to note that the
projectors $Q$ and $R$ are both linear and can be recorded for the
fulfillment of projection operations in the particular Euclidean
space. The projection operators $Q$ and $R$ allow one to carry out
the splitting of Euclidian space of vectors $B$, where $B(0)$ and
$B(t) \in B$, into a straight sum of the two mutually
supplementary subspaces in the  following way: \be B=B' + B'',\
B'=QB,\ B''=RB. \label{rule_of_com} \ee

Let us consider the finite-difference Liouville's Eq.
\Ref{eq_of_m} for the vector of fluctuations of a Brownian
particle position \be \frac{\triangle}{\triangle t}
\textbf{B}_{m+k}^{m}(t)=i\hat{\mathcal{L}}(t,\tau)\textbf{B}_{m+k}^{m}(t).
\label{Eq_of_m_B} \ee Affecting the last equation by the operators
$Q$ and $R$ successfully, we can split the dynamic equation
\Ref{Eq_of_m_B} into two interconnected equations in the two
mutually-supplementary Euclidean subspaces: \bn
\frac{\triangle}{\triangle t}
\textbf{B}'(t)=iQ\hat{\mathcal{L}}(Q+R)\textbf{B}(t)=i\hat{\mathcal{L}}_{11}\textbf{B}'(t)
+i\hat{\mathcal{L}}_{12}\textbf{B}''(t), \label{first}\\
\frac{\triangle}{\triangle t}
\textbf{B}''(t)=iR\hat{\mathcal{L}}(Q+R)\textbf{B}(t)=i\hat{\mathcal{L}}_{21}\textbf{B}'(t)
+i\hat{\mathcal{L}}_{22}\textbf{B}''(t). \label{second} \en In the
above equations the matrix elements $\hat{\mathcal{L}}_{ij}$ of
the quasi-operator $\hat{\mathcal{L}}$ have been introduced \bn
\hat{\mathcal{L}}_{11}=Q\hat{\mathcal{L}}Q,\
\hat{\mathcal{L}}_{12}=Q\hat{\mathcal{L}}R,\
\hat{\mathcal{L}}_{21}=R\hat{\mathcal{L}}Q,\
\hat{\mathcal{L}}_{22}=R\hat{\mathcal{L}}R, \nonumber \en \bn
\hat{\mathcal{L}}=
\begin{pmatrix}
  \hat{\mathcal{L}}_{11} & \hat{\mathcal{L}}_{12} \\
  \hat{\mathcal{L}}_{21} & \hat{\mathcal{L}}_{22}
\end{pmatrix}.
\en
Operators $\hat{\mathcal{L}}_{ij}$ act as follows:
$\hat{\mathcal{L}}_{11}$ - from a subspaces $B'$ to $B'$;
$\hat{\mathcal{L}}_{12}$ - from $B''$ to $B'$;
$\hat{\mathcal{L}}_{21}$ - from $B'$ to $B''$; and
$\hat{\mathcal{L}}_{22}$ - from $B''$ to $B''$.

To simplify the Liouville Eqs. \Ref{first} and \Ref{second}, we
exclude the irrelevant part $B''(t)$ and construct the closed
equation for the relevant part $B'(t)$. For this purpose, it is
necessary to obtain a step-by-step solution of Eq. \Ref{second}
\bn \frac{\triangle \textbf{B}''(t)}{\triangle
t}&=&\frac{\textbf{B}''(t+\tau)-\textbf{B}''(t)}{\tau}=i\hat{\mathcal{L}}_{21}
\textbf{B}'(t)+i\hat{\mathcal{L}}_{22}\textbf{B}''(t),\\
\textbf{B}''(t+\tau)&=&\textbf{B}''(t)+i\tau
\hat{\mathcal{L}}_{21}\textbf{B}'(t)+i\tau\hat{\mathcal{L}}_{22}\textbf{B}''(t)\nonumber\\
&=&(1+i\tau\hat{\mathcal{L}}_{22})\textbf{B}''(t)+i\tau\hat{\mathcal{L}}_{21}\textbf{B}'(t)\nonumber\\
&=&U_{22}(t+\tau,t)\textbf{B}''(t)+i\tau\hat{\mathcal{L}}_{21}(t+\tau,t)\textbf{B}'(t),
\label{irrelevant} \en where
$U_{22}(t+\tau,t)=1+i\tau\hat{\mathcal{L}}_{22}(t+\tau,t)$ is the
operator of a time step shift.

With the help of Eq. \Ref{irrelevant} we can derive the following
expression for the next time step: \bn
\textbf{B}''(t+2\tau)&=&U_{22}(t+2\tau,t+\tau)\textbf{B}''(t+\tau)
+i\tau\hat{\mathcal{L}}_{21}(t+2\tau,t+\tau)\textbf{B}'(t+\tau)\nonumber\\
&=& U_{22}(t+2\tau,t+\tau)[U_{22}(t+\tau,t) \textbf{B}''(t)+
i\tau\hat{\mathcal{L}}_{21}(t+\tau,t) \textbf{B}'(t)]\nonumber\\
& &+i\tau\hat{\mathcal{L}}_{21}(t+2\tau,t+\tau)
\textbf{B}'(t+\tau)\nonumber\\
&=& U_{22}(t+2\tau,t+\tau)U_{22}(t+\tau,t) \textbf{B}''(t) + i\tau
[U_{22}(t+2\tau,t+\tau)\nonumber\\
& &\times \hat{\mathcal{L}}_{21}(t+\tau,t)\textbf{B}'(t) +
\hat{\mathcal{L}}_{21}(t+2\tau,t+\tau) \textbf{B}'(t+\tau)]. \en

Generalizing this result in case of the $m$th discrete step we
find the following final result \bn \textbf{B}''(t+m\tau) &=&
\left \{ \hat{T} \prod_{j=0}^{m-1}
U_{22}(t+(j+1)\tau,t+j\tau)  \right \} \textbf{B}''(t) \nonumber\\
& & +i\tau \sum_{j=0}^{m-1} \left \{ \hat{T} \prod_{j'=j}^{m-2}
U_{22}(t+(j'+2)\tau,t+(j'+1)\tau) \right \} \nonumber\\
& & \times \hat{\mathcal{L}}(t+(j+1)\tau, t+j\tau)
\textbf{B}'(t+j\tau). \label{evol} \en

Here $\hat{T}$ denotes the Dyson operator of chronological
ordering. Substituting the irrelevant part of Eq. \Ref{first} in
the right side of Eq. \Ref{evol}, we obtain the closed
finite-difference equation for the relevant part of the
correlation state vector \bn \frac{\triangle}{\triangle t}
\textbf{B}'(t+m\tau)&=& i \hat{\mathcal{L}}_{11}(t +(m+1)\tau,
t+m\tau) \textbf{B}'(t+m\tau)
+i\hat{\mathcal{L}}_{12}(t+(m+1)\tau,t+m\tau)\nonumber\\
& & \times  ( \left \{ \hat{T}\prod_{j=0}^{m-1}
U_{22}(t+(j+1)\tau,t+j\tau) \right\} \textbf{B}''(t) - \tau
\sum_{j=0}^{m-1} \{ \hat{T}\prod_{j'=j}^{m-2}
U_{22}(t+(j'+2)\tau,t\nonumber\\
& & +(j'+1)\tau) \} \hat{\mathcal{L}}_{21}(t+(j+1)\tau,t+j\tau)
\textbf{B}'(t+j\tau) ). \label{evol_two} \en

Substituting Eqs. \Ref{oper} and \Ref{rule_of_com} in Eq.
\Ref{evol_two}, we derive a finite-difference kinetic equation of
a non-Markov type for TCF $b(t)$

\bn \frac{\triangle b(t)}{\triangle t}= \lambda_{1}b(t) - \tau
\Lambda_{1} \sum_{j=0}^{m-1}M_{1}(t-j\tau)b(j\tau).
\label{first_equation} \en Here the TCF $M_{1}(t)$ is the first
order memory function \be M_{1}(t-j\tau)=\frac{ \langle
\textbf{W}_{1}(0)  \textbf{W}_{1}(t-j\tau)
\rangle}{|\textbf{W}_{1}(0)|^{2}}, \label{first_MF} \ee Here
$\Lambda_{1}$ is the frequency relaxation parameter of the first
order with a square frequency dimension, and is defined by Eq.
\Ref{noise_parameters}. Then if the dynamic variable
$\textbf{W}_{0}= \textbf{B}_{k}^{0}(0)$ represents the
fluctuations of a Brownian particle position, the dynamic variable
$\textbf{W}_{1}=i\hat{\mathcal{L}}\textbf{W}_{0}-
\lambda_{1}\textbf{W}_{0}$ contains fluctuations of pulses of a
Brownian particle [see Eqs. \Ref{Dynamic_var}]. The function
$M_{1}(t)$ is time correlation function of fluctuations of a
Brownian particle velocity.

Defining the corresponding projection operators to new dynamic
variable $\textbf{W}_{1}$ and repeating the above described
procedure, we find the finite-difference kinetic equation for
$M_{1}(t)$ \bn \frac{\triangle M_{1}(t)}{\triangle t}=
\lambda_{2}M_{1}(t) - \tau \Lambda_{2}
\sum_{j=0}^{m-1}M_{2}(t-j\tau)M_{1}(j\tau).
\label{second_equation} \en Here $\lambda_{2}$ and $\Lambda_{2}$
are the frequency relaxation parameters of the second order,
$M_{2}(t)$ is the second order memory function or memory function
of the velocity correlation function for a Brownian particle
(memory friction) \cite{Tuckerman,Berne,Hanggi,Hynes}. In fact,
Eq. \Ref{second_equation} is a discrete finite-difference
\textit{generalized Langevin equation} (GLE). So, $M_{2}(t)$ can
be associated  with TCF of the stochastic Langevin forces, for
which the similar equation can be received \bn \frac{\triangle
M_{2}(t)}{\triangle t}= \lambda_{3}M_{2}(t) - \tau \Lambda_{3}
\sum_{j=0}^{m-1}M_{3}(t-j\tau)M_{2}(j\tau) \label{third_equation}
\en with the third order frequency relaxation parameters
$\lambda_{3}$ and $\Lambda_{3}$, and the memory function of the
third order $M_{3}(t)$ respectively.

The three Eqs. \Ref{first_equation}, \Ref{second_equation} and
\Ref{third_equation} are the exact consequence of microscopic
discrete finite-difference equations of motion. Calculation of the
memory function $M_{i}(t)$ and the relaxation frequency parameters
$\lambda_{j}$, $\Lambda_{k}$ are the central point here.

In case of a Brownian particle presented above, dynamic variables
$\textbf{W}_{0}$ and $\textbf{W}_{1}$ are the position and pulses
of random particles. In fact, they can be taken as characteristics
of any other non-stationary process. The averaging operator of
sampling with the length $M$, $\hat{A}=1/M\sum_{j=0}^{M-1}$ [see
Eq. \Ref{average_1}], applied to the intermediate local sampling,
can be changed by any other operator, depending on the goal of the
investigation. The operator $\hat{A}$ allows one to get clear of
sharp fluctuations in the initial sampling of data $X(t)$ and to
replace it by an other $Y(t')$: $X(t) \rightarrow Y(t')$, which
contains the results of coarse-graining averaging. However, it is
not always convenient to average a local sampling. In these cases
the operator $\hat{A}$ can be replaced by an other operator, which
allows one to obtain only one number from every local sampling. In
general case, an another, a more universal method of
transformation of the initial sampling into the sampling with some
specified (required) characteristics can become discrete
wavelet-transform \cite{Daub,Wave_1,Wave_2,Wave_3,Astafieva},
which is defined by the following equation: \be
W(j,k)=\sum_{j}\sum_{k} X(k)2^{-j-2}\Psi(2^{-j}n-k). \label{dwt}
\ee Here $X(k)$ is the sampling \Ref{series} and $\Psi(t)$ is a
time function with fast decay called mother wavelet. The following
analysis can be applied to the new transformed data $W(j,k)$
according to the algorithm described above. Namely, memory
functions $M_{i}(t)$ and frequency relaxation parameters
$\lambda_{i}$ and $\Lambda_{i}$ can be calculated for the
transformed data.

%%%%%%%%%%%%%%%%%%%%%%%%%%%%%%%%%%%%%%%%%%%%%%%%%%%%%%%%%%

\subsubsection{Analogue of Green-Kubo relation for diffusion
coefficient for time discrete system}

%%%%%%%%%%%%%%%%%%%%%%%%%%%%%%%%%%%%%%%%%%%%%%%%%%%%%%%%%%%

According to the theory of random walkers the mean-square
displacements of a Brownian particle can be defined as \be \langle
\triangle y^{2} \rangle_{t}= \int_{-\infty}^{+\infty} \triangle
y^{2} \Phi_{1}(y,t)dy=2Dt, \hspace{1.5cm} (t \to \infty).
\label{Einst} \ee Here $\Phi_{1}(y,t)$ is  the density of
probability of being of particle at point $y$ at time moment $t$.
The Eq. \Ref{Einst} is a well known Einstein relation for
continuous displacements.

According to Eq. \Ref{Einst} the displacement during time $t$ is
\be \triangle y_{t}=y(t)-y(0). \label{diff} \ee

In case of the discrete time Eq. \Ref{diff} can be rewritten in
the form \be \triangle y(t)=\triangle
y(T+j\tau)=y(T+(j+m)\tau)-y(T+j\tau), \ t=m\tau,\hspace{1.5cm}
m\gg 1. \ee Then the mean-square displacement of a Brownian
particle is \be \langle \triangle
y^{2}(T+(j+m)\tau)\rangle=\frac{1}{N-m}
\sum_{j=0}^{N-m-1}[y(T+(j+m)\tau)-y(T+j\tau)]^{2}, \hspace{1.5cm}
N>m\gg1, \ee where $N-m$ is the quantity of ''possible ways''. The
diffusion coefficient takes the following form \be
D=\frac{1}{2m(N-m)\tau}\sum_{j=0}^{N-m-1}[y(T+(j+m)\tau)-y(T+j\tau)]^{2},
\hspace{1.5cm} \textrm{at}\ N >m\gg 1\ (\textrm{or}\ t \to \infty\
). \label{diff_2} \ee Let's consider separately the sum in the
last expression in terms of the velocity [see Eq. \Ref{velocity}]
\bn
\sum_{j=0}^{N-m-1}[&y&(T+(j+m)\tau)-y(T+j\tau)]^{2}\nonumber\\
&=&\underbrace{[y(T+m\tau)-y(T)]^{2}
+[y(T+(1+m)\tau)-y(T+\tau)]^2+\ldots+ [y(T+(N-1)\tau)-
y(T+(N-m-1)\tau)]^{2}}_{(N-m)\
\textrm{of square brackets}\ [\ldots]}\nonumber\\
&=&[y(T+m\tau)-y(T+(m-1)\tau)+y(T+(m-1)\tau)-...-y(T)]^{2}+
\ldots \nonumber\\
& &[y(T+(N-1)\tau)-y(T+(N-2)\tau)+y(T+(N-2)\tau)-...
-y(T+(N-m-1)\tau)]^{2}\nonumber\\
&=&\tau^{2}[v(T+(m-1)\tau)+v(T+(m-2)\tau)+...+v(T)]^{2}+\ldots\nonumber\\
&+&
\tau^{2}[v(T+(N-2)\tau)+v(T+(N-3)\tau)+\ldots+v(T+(N-m-1)\tau)]^{2} \nonumber\\
&=&\tau^{2}\sum_{j=0}^{N-m-1} \left [ \sum_{j=0}^{m-1+j}
v(T+k\tau) \right ]^{2}. \label{promezh} \en Then the expression
for the diffusion coefficient \Ref{diff_2} can be rewritten as:
\be D=\frac{\tau}{2m(N-m)} \sum_{j=0}^{N-m-1} \left [
\sum_{k=j}^{m-1+j} v(T+k\tau) \right ]^{2}. \label{Green_Kubo} \ee

Eq. \Ref{Green_Kubo} is the discrete finite-difference analog of
the famous Green-Kubo relation for the diffusion coefficient.
Given relation has been obtained from Einstein equation
\Ref{Einst} for a discrete system. The asymptotic limit $t \to
\infty$ can be replaced here by the similar condition $N\to
\infty$, $m \to \infty$, $N>m$.

%%%%%%%%%%%%%%%%%%%%%%%%%%%%%%%%%%%%%%%%%%%%%%%%%%%%%%%
\section{Local noisy parameters} \label{local_parameters}
%%%%%%%%%%%%%%%%%%%%%%%%%%%%%%%%%%%%%%%%%%%%%%%%%%%%%%%
Seismic data represent discrete random series, which is recording
of displacements of the Earth's surface. Therefore, we can use the
above-stated formalism to analyze seismic data. In particular, the
local dependence of various characteristics
\cite{Stanley,Yulmetyev1} can be serve as additional source of
information on properties of objects. The noise parameters
$\lambda_{i}$ and $\Lambda_{i}$ are very sensitive to the presence
of a nonrandom component in the sampling. The change of the
character of the correlated noise and, the appearance of the
additional signal in the sampling can cause the alternation of
these parameters. So, the time behavior of the local parameters
$\lambda_{i}$, $\Lambda_{i}$ is important and informative for
analyzing of seismic data.

The procedure of localization consists in the following. Let us
assume that we have an array of data $\{ x_{1},\ \ x_{2},\ \
x_{3},..., x_{M},...,x_{N} \}$ and take the initial sampling of
the fixed length $M$. Then by passing through all array of values
with the "work window" of the fixed length $M$ we can calculated
the time series of the noisy parameters $\{
\lambda_{i}(T,T+M\tau),\ \
\lambda_{i}(T+\tau,T+(M+1)\tau),...,\lambda_{i}(T+(N-M-1)\tau,T+(N-1)\tau)
\}$ and $\{ \Lambda_{i}(T,T+M\tau),\ \
\Lambda_{i}(T+\tau,T+(M+1)\tau),...,\Lambda_{i}(T+(N-M-1)\tau,T+(N-1)\tau)
\}$

Obviously, it is inadmissible to use both very large intervals
($M=$1000 and more points) and very short intervals ($M=$50 points
and less) for definition of local parameters $\lambda_{i}(t)$,
$\Lambda_{i}(t)$. In the first case the physical sense of the
localization procedure is lost. On the other hand, it is
impossible to carry out any plausible correlation analysis with
small intervals because of gross errors. Therefore there is
necessity for finding the optimum length of the initial local
interval or quantity $M$.

To determine the optimal minimal local sampling we have used the
data corresponding to the calm state of the Earth before the
technogenic explosion (an underground nuclear explosion). The
calculation procedure consists in the following. We have taken the
interval of $40$ points as the starting point and have calculated
all the low-order noise parameters $\lambda_{i}$ $(i=1,2,3)$ and
$\Lambda_{j}$ $(j=1,2)$ by Eqs. \Ref{noise_parameters} and
\Ref{Dynamic_var}. Then the interval was consistently increased by
unit time segment $\tau$ and the relaxation parameters
$\lambda_{i}$, $\Lambda_{j}$ were calculated every time at the
increase of the interval. As a result of this procedure executed
for the calm state of the Earth we have established that all
parameters take "steady" numerical values at the interval
approximately equal to $150 $ points and more (see Fig. 1 for more
details). Namely, from Fig. 1 one can see, that the parameters
$\lambda_{1}$ and $\Lambda_{1}$ take minimal values in their
absolute quantity at this length of interval, the amplitude of
value fluctuations of the parameter $\Lambda_{2}$ gets lower and
level off. It is again the evidence of reduction of the noise
influence. Fluctuations of the parameters $\lambda_{2}$ and
$\lambda_{3}$ also decrease, and the parameters themselves take
steady values starting with the sampling of length $\sim 150$
points. Applying this procedure to other data of the calm state of
the Earth, we find the same behavior of noise parameters
$\lambda_{i}$, $\Lambda_{i}$ and detect the minimal interval of
$150$ points again. So, we choose the interval of such length as
being optimal for accumulation of local statistics \cite{Herst}.

%%%%%%%%%%%%%%%%%%%%%%%%%%%%%%%%%%%%%%%%%%%%%%%%%%%%%%%%%%%%%%%%%%%%%
\section{Definition of relaxation parameters $\lambda_{i}(t)$ and $\Lambda_{j}(t)$ for earthquakes
and technogenic explosions data} \label{definition of local
parameters}
%%%%%%%%%%%%%%%%%%%%%%%%%%%%%%%%%%%%%%%%%%%%%%%%%%%%%%%%%%%%%%%%%%%%%

It is well known that modern seismic devices allow one to derive
different quantitative and qualitative data about the seismic
state of the Earth. In this work we analyse three various weak
local earthquakes (EQ's) in Jordan (1998) [EQ(1), EQ(2), EQ(3)],
one strong earthquake in Turkey (summer, 1999) [EQ(T)] and three
local underground technogenic explosions (TE) [TE(1), TE(2),
TE(3)] with the length of registration from $10000$ to $25000$
points. In case of strong EQ its seismogram contains $65000$
registered points. All these experimental data were courteously
given by the Laboratory of Geophysics and Seismology (Amman,
Jordan). All data correspond to transverse seismic displacements.
The real temporal step of digitization $\tau$ between the
registered points of seismic activity has the following values,
viz, $\tau=0.02s$ for the EQ(T), and $\tau=0.01s$ for all others
cases.
\begin{table}
\caption{Some characteristics of Technogenic Explosions [TE(1),
TE(2), TE(3)] and Earthquakes [EQ(1), EQ(2), EQ(3), EQ(T)]
obtained from seismic data: $a_{1}$ is the maximal amplitude of
signal oscillations before the ''event'', $a_{2}$ is the maximal
amplitude of signal oscillations at the time of the ''event'',
$T_{0}$ is the time from the beginning of the signal registration
to the beginning the ''event'', $T_{l}$ is the ''event'' duration,
$T_{total}=N$ is the total time of signal recording.}
   \begin{center}
    \begin{tabular}{l|cc|c|ccc}
    \hline
    \hline
    &  $a_{1}$  &  $a_{2}$  &  $a_{2}/a_{1}$  &  $T_{0}$  &  $T_{l}$  &  $T_{total}$  \\ %$\lambda_{0}$ & $\triangle \lambda $ & $T_{\lambda}$ & $\Lambda_{0}$ & $\triangle \Lambda $ & $T_{\Lambda}$ &  & $T_{l}/T_{\lambda}$ \\
    \hline TE(1) & $1.03 \cdot 10^{-3}$ & $8.24 \cdot 10^{-3}$ & 8 & 4091 & 4500 & 12500\\
           TE(2) & $1.5 \cdot 10^{-3}$ & $18.5 \cdot 10^{-3}$ & 12.3(3) & 3538 & 4500 & 10000\\
           TE(3) & $1.18 \cdot 10^{-3}$ & $56.5 \cdot 10^{-3}$ & 47.88 & 4091 & 4850 & 15000\\
           EQ(1) & $0.74 \cdot 10^{-3}$ & $7.1 \cdot 10^{-3}$ & 9.59 & 5682 & 4500 & 15000 \\
           EQ(2) & $0.94 \cdot 10^{-3}$ & $6.1 \cdot 10^{-3}$ & 6.49 & 12308 & 5770 & 25000\\
           EQ(3) & $0.59 \cdot 10^{-3}$ & $14.1 \cdot 10^{-3}$ & 23.9 & 3182 & 5700 & 12500\\
           EQ(T) & $9 \cdot 10^{-3}$ & $20$ & $2.2(2) \cdot
           10^{3}$ & $\approx 13500$ & -
           & $65000$\\
           \hline
           \hline
    \end{tabular}
   \end{center}
   \label{table_1}
\end{table}
We defined such characteristics as the maximal amplitude of signal
fluctuations before ''event'' $a_{1}$, the maximal amplitude of
signal fluctuations during ''event'' $a_{2}$, EQ (TE) power
$a_{2}/a_{1}$, time interval till EQ (TE) $T_{0}$, continuance of
''event'' $T_{l}$ and finally the total time of signal recording
$T_{total}$ directly from seismic data. These quantities are
presented in Table I. They give a clear notion about the duration
and power of investigated phenomena. One can see from Table I,
approximately $4500-5700$ points are accounted for the visible
part a wavelet. This number of the recorded points allows one to
execute analysis by means of the local parameters $\lambda_{i}(t)$
and $\Lambda_{j}(t)$.

The procedure of calculation of time-dependence for
$\lambda_{i}(t)$ and $\Lambda_{j}(t)$ was based on the following
operations. An interval with $M \sim 150$ points is taken, and
noisy parameters $\lambda_{i}$, $\Lambda_{j}$ are calculated for
it with the help of Eqs. \Ref{noise_parameters},
\Ref{Dynamic_var}. Then the operation of ''stepwise shift to the
right'' at the interval of the fixed length $M$ is executed, and
parameters are computed again. These actions are executed, while
the initial sampling $X(t)$ will not be finished.  As a result we
obtain the following dependencies $\{ \lambda_{i}(T,T+M\tau),\ \
\lambda_{i}(T+\tau,T+(M+1)\tau),...,\lambda_{i}(T+(N-M-1)\tau,T+(N-1)\tau)
\}$ and $\{ \Lambda_{i}(T,T+M\tau),\ \
\Lambda_{i}(T+\tau,T+(M+1)\tau),...,\Lambda_{i}(T+(N-M-1)\tau,T+(N-1)\tau)
\}$. If the character of the noise in the investigated data
change, some signal will appear or disappear, and it will be
directly reflected in the behavior of the relaxation
characteristics.

The results of the above described procedure for the case of EQ
and TE data are shown in Figs. $2$, $3$ and Table II. However, in
order to check up the optimized length of the local interval, we
calculated local parameters $\lambda_{i}(t)$ and $\Lambda_{j}(t)$
$(i=1,2,3)$, $(j=1,2)$ at the local sampling with the length
$M=100,200,250,300,350$ and $400$ points. It turned out that a
large number of various noises in the behavior of $\lambda_{i}(t)$
and $\Lambda_{j}(t)$ are superimposed on the carrying trajectory
at $M=100$ points (parameters has a gross errors). The line-shapes
of $\lambda_{i}(t)$ and $\Lambda_{j}(t)$ practically cease to
change at the sampling the length $M=150$ and more,
$M=200,250,...$. Once again it is evidence that the local interval
with the length $M=150$ points is optimal for the analysis of
strong, weak EQ's and TE's.

The detailed analysis of the received results allows one to reveal
the following features.

\textbf{Weak EQ's and  local underground TE's} (for TE(3) and
EQ(3), see Figs. 2, 3):

\textbf{1.} All the parameters $\lambda_{i}(t)$ take only negative
values ($\lambda_{i}(t) < 0$), whereas the noisy parameters
$\Lambda_{j}(t)$ can be both positive and negative.

\textbf{2.} \textit{Noisy parameter $\lambda_{1}(t)$.} The
absolute magnitude $|\lambda_{1}|$ increases sharply in its
amplitude by a factor approximately equal to $4-13.3$ during EQ
(various for different EQ's), and then it returns to its initial
state. the restoration time $T_{\lambda_{1}}$ and the duration of
''event'' $T_{l}$ are approximately equal for weak EQ, i. e.
$T_{\lambda_{1}} \approx T_{l}$.

The absolute magnitude $|\lambda_{1}|$ also exhibits an abrupt
rise $\sim 2.2-3.5$ times higher for TE's. However, it returns
quickly to its normal level. The restoration time $\lambda_{1}(t)$
for TE is less than the duration of the ''event'' approximately by
a factor of $2.5-3$.

\textbf{3.} \textit{Noisy parameter $\lambda_{2}(t)$.} The
parameter $\lambda_{2}(t)$ responds to the beginning of the
''event'' by an abrupt rise of its value. It increases in
amplitude both for TE's and for EQ's. The character of the noise
changes during the ''event''. The parameter responds to the power
of the ''event''.

\textbf{4.} \textit{Noisy parameter $\lambda_{3}(t)$.} This
parameter always fluctuates near its numerical value $-1$, and has
an abrupt rise at the beginning the ''event'' in the form of
separated spikes (see Figs. 3 and 4). The parameter keenly
responds to the noise changes.

\textbf{5.} \textit{Noisy parameter $\Lambda_{1}(t)$.} It
fluctuates near zero before and after the ''event'' changing its
sign at this time. The parameter increases sharply at the
''event'' and \textit{always} \textbf{(!)} retains positive. Then
it decays smoothly. Restoration time $T_{\Lambda_{1}}$ and the
duration of the ''event'' $T_{l}$ are approximately the same both
for the EQ and TE. This parameter is very sensitive to the changes
of the noise character. For example, the data analysis of the
EQ$(2)$ shows that the separate burst of the amplitude values of
$\Lambda_{1}(t)$ appears for $\approx 4000$ points up to the
beginning of the EQ, although such indicator was not visually
observed in the initial seismic data \cite{param}. It may be the
evidence of high prognostic property of this parameter for EQ's
forecasting.

\textbf{6.} \textit{Noisy parameter $\Lambda_{2}(t)$.} Noise
changes of the parameter $\Lambda_{2}(t)$ are observed during the
''event'' both during EQ's and TE's. From Figs. $2$ and $3$ we can
see that $\Lambda_{2}(t)$ has a distinctive negative depression
during the ''event''.

\textbf{Strong EQ's} (for EQ(T) see Fig. 4) \cite{Turkish}:

All parameters react keenly to the appearance of the signal in
case of a strong EQ. Let us consider the behavior of the local
parameters in this case in detail.

\textbf{7.} \textit{Noisy parameter $\lambda_{1}(t)$.} The
parameter demonstrates the presence of noise and takes negative
values before the ''event'' [Region $I$ in Fig. 4 f)]. As the
''event'' approaches, the amplitude of fluctuations decreases, and
oscillations turn into negligible fluctuations near zero value for
II, III and IV Regions (for more detail, see Fig. 4).

\textbf{8.} \textit{Noisy parameter $\lambda_{2}(t)$.} A noise is
also observed in the behavior of this parameter before the
''event'', and this parameter takes only negative values. However,
its values decrease sharply in absolute magnitude and begin to
take values near zero with the appearance of an EQ signal (Fig. 4
b).

\textbf{9.} \textit{Noisy parameter $\lambda_{3}(t)$.} This
parameter takes only negative values at all times $t$. The
negligible noise appears before the ''event'. As the ''event''
approaches the noise increases in amplitude by the factor $3-5$.
The amplitude of oscillations begins to decrease in Region IV (see
Fig. 4 c).

\textbf{10.} \textit{Noisy parameters $\Lambda_{1}(t)$ and
$\Lambda_{2}(t)$.} Parameters fluctuate, taking positive values
mainly before the ''event'' (Region I). With the beginning of
''event'' values of parameter increase greatly in absolute
magnitude approaching zero. At the beginning of Region II  both
parameters change their sign from positive to negative. Then for
Regions II, III and IV we see only right line with
$\Lambda_{1}(t), \Lambda_{2}(t)\rightarrow 0$ on the scales of
Figs. 4 d) and e). However, the parameter $\Lambda_{2}(t)$ has
faintly visible fluctuations for Region IV characterizing the
final EQ phase [see Figs. 4 d) and e)].

So, all parameters are very sensitive to the approach of a strong
EQ. A sharp change in their behavior is appreciable before strong
fluctuations of the Earth's surface far off $10000$ points ($\sim
3.5$ minutes).

%%%%%%%%%%%%%%%%%%%%%%%%%%%%%%%%%%%%%%%%%%%%%%%%%%%%%%%%%%%%%
\section{Simple exponential model for time local behavior of
noisy relaxation parameters $\lambda_{1}(t)$ and $\Lambda_{1}(t)$}
\label{model}
%%%%%%%%%%%%%%%%%%%%%%%%%%%%%%%%%%%%%%%%%%%%%%%%%%%%%%%%%%%%%

As can be seen from Figs. $2-4$ all relaxation parameters are very
sensitive to the beginning of EQ and TE. The behavior of
parameters $\lambda_{1}(t)$ and $\Lambda_{1}(t)$ in weak local
EQ's and local TE's is of great interest. These parameters are
initial in our calculations [see Eqs. \Ref{noise_parameters},
\Ref{Dynamic_var}]. The analysis has shown that parameters
oscillate near average values $\lambda_{0}$ and $\Lambda_{0}$,
correspondingly, before and after the oscillations visible on
seismograms. The results of the behavior of these parameters for
TE(3) and EQ(3) are presented in Figs. $2$ a)-d) and Fig. $3$ a)-
d). However, a sudden rise by factors $\Delta\lambda_{0}$ and
$\Delta\Lambda_{0}$ is always observed in the behavior of these
parameters at the enhancement or the appearance of the signal (it
is seen at the beginning of the EQ or the TE in seismogram data).
Furthermore, the continuous attenuation occurs. Over all this time
these parameters have a well-defined pronounced trend. Such
behavior of $\lambda_{1}(t)$ and $\Lambda_{1}(t)$ give us a
possibility of modelling the time dependence of these parameters
by some simple mathematical functions.
\begin{table}
\caption{Characteristics of local parameters
$\lambda_{1}(t)$ and $\Lambda_{1}(t)$: $T_{\lambda_{1}}$ is the
parameter $\lambda_{1}(t)$ relaxation time, $T_{\Lambda_{1}}$ is
the relaxation time of $\Lambda_{1}(t)$, $\tau_{\lambda}$ is the
relaxation time for exponential attenuation of $\lambda_{1}(t)$,
$\tau_{\Lambda}$ is the relaxation time for exponential
attenuation of $\Lambda_{1}(t)$.}
   \begin{center}
    \begin{tabular}{l|c|c|c|c|c|c|c|c|c|c|c}
    \hline
    \hline
    & $\lambda_{0}$ (units of $\tau^{-1}$) & $\triangle \lambda $ (units of $\tau^{-1}$) & $T_{\lambda}$ & $\tau_{\lambda}$ & $\Lambda_{0}$ (units of $\tau^{-2}$) & $\triangle \Lambda $ (units of $\tau^{-2}$) & $T_{\Lambda}$ & $\tau_{\Lambda}$ & $\Delta \lambda/\lambda_{0}$ & $\Delta\Lambda/\Lambda_{0}$ & $T_{l}/T_{\lambda}$\\
    \hline TE(1) & -0.13 & -0.425 & 1800 & 90 & 0.02 & 0.28 & 4500 & 45 & 3.27 & 14 & 2.5\\
           TE(2) & -0.15 & -0.34 & 1440 & 100 & 0.02 & 0.28 & 4032 & 55  & 2.26(6) & 14 & 3.125\\
           TE(3) & -0.15 & -0.33 & 1870 & 80 & 0.002 & 0.42 & 4850 & 45 & 2.2 & 210 & 2.59\\
           EQ(1) & -0.045 & -0.6 & 4500 & 170 & 0.005 & 0.43 & 4500 & 110 & 13.3(3) & 86 & 1\\
           EQ(2) & -0.1 & -0.5 & 5770 & 170 & 0.001 & 0.35 & 5770 & 130 & 5 & 350 & 1\\
           EQ(3) & -0.12 & -0.47 & 5700 & 210 & 0.01 & 0.35 & 5700 & 180 & 3.91(6) & 35 & 1\\
           \hline
           \hline
    \end{tabular}
   \end{center}
   \label{tab_2}
\end{table}
The fitting procedure showed that the time behavior of these
parameters can be well approximated by the following simple time
dependencies: \be \lambda_{1}(t)=\lambda_{0}+\Delta\lambda\cdot
\textrm{exp}\left
\lbrace-\frac{t-T_{0}}{T_{\lambda}}\right\rbrace\cdot H(t-T_{0}),
\label{expon_1} \ee \be
\Lambda_{1}(t)=\Lambda_{0}+\Delta\Lambda\cdot \textrm{exp}\left
\lbrace-\frac{t-T_{0}}{T_{\Lambda}}\right\rbrace\cdot H(t-T_{0}),
\label{expon_2} \ee where $H(t)$ is the Heaviside function,
$T_{\lambda}$ and $T_{\Lambda}$ are the relaxation times of
$\lambda_{1}(t)$ and $\Lambda_{1}(t)$, correspondingly. The time
$T_{0}$ is the same in Eqs. \Ref{expon_1} and \Ref{expon_2} for
parameters $\lambda_{1}(t)$ and $\Lambda_{1}(t)$ (see Table
\ref{table_1}). The numerical values of the variables, included in
Eqs. \Ref{expon_1}, \Ref{expon_2} were defined for EQ's and TE's
by comparison of localization results with these equations (see
Fig. $5$). Numerical values of parameters are presented in Table
\ref{tab_2}.

So, it proved that the restoration of these parameters to their
steady values occurs according to the \textit{exponential} law. As
can be seen from Fig. $5$ this description best suits for
$\lambda_{1}(t)$ and $\Lambda_{1}(t)$ of weak EQ's. The foregoing
estimations strengthen fully our resume \textbf{2} and \textbf{5}
of Section \ref{definition of local parameters}.

The results presented in the last three columns of Table II might
be of interest for readers. As the quantities $\triangle
\lambda_{0}/\lambda_{0}$ and $\triangle \Lambda_{0}/\Lambda_{0}$
show the rise of value of the corresponding parameter at TE and
EQ. Finally, the ratio between the "event"\ duration $T_{l}$ and
the relaxation time $T_{\lambda}$ discovers a remarkable
distinction between TE's and  weak EQ's.

%%%%%%%%%%%%%%%%%%%%%%%
\section{Conclusion} \label{conclusion}
%%%%%%%%%%%%%%%%%%%%%%%

In this work universal method for investigating non-stationary,
unsteady and non-Markov random processes in discrete systems is
suggested. This universality is achieved by combining the
opportunities of microscopic, mesoscopic and macroscopic
descriptions for complex systems. This method allows one to find
and to analyze fast, slow and super-slow processes. To investigate
super-slow processes we propose to use the model of a
"quasi-Brownian particle". The wavelet-transformation of the
initial time series can be used for this purpose. This method
helps to analyze and differentiate similar signals of different
origin. Theoretical investigations have been realized by means of
two methods supplementing each other: the statistical theory of
discrete non-Markov stochastic processes \cite{Yulmetyev1} and the
local noisy parameters. The application of the last method gives a
possibility to study non-stationary and unsteady processes with
alternation and superimposition of different modes. The
correlation between the time scales characteristic of different
modes may be different. However, at accurate  realization the
localization procedure allows one to separate the noise and the
signals \cite{Yulmetyev1,Stanley}, and to carry out their
quantitative and informative analysis.

Another important advantage of this method is the possibility to
operate it in ''real time''\ regime, i.e. it can be put into
practice immediately at getting the data, that is of great
practical value.

The developed approach has been tested for strong and weak EQ's
data and nuclear underground TE's. As a result we have obtained of
the following results.

The time-behavior of the local relaxation parameters can be
described by simple model relaxation functions. The temporal
relaxation of parameters $\lambda_{1}(t)$ and $\Lambda_{1}(t)$ in
weak EQ's and TE's after the beginning of the ''event'' occurs
according to the exponential law. However, the restoration and
duration of the events are practically the same in case of weak
EQ's. The restoration time of parameters $\lambda_{1}(t)$ and
$\Lambda_{1}(t)$ in the case of TE's differs noticeably from the
duration of the event. So, this approach can be useful in
recognition these two different seismic phenomena. From the
analysis of strong EQ data one can see that the behavior of all
parameters changes greatly long before the EQ. For example, such
change for the EQ(T) presented here occurs $\sim 3.5$ minutes
before the main event. This change of $\lambda_{i}(t)$ and
$\Lambda_{i}(t)$ obtained for strong EQ's opens the possibility
for a more accurate registration of the beginning of changes of
the parameters before the visual wavelet and real EQ's.

We are sure that the suggested method can be very useful for the
study of a wide class of random discrete processes in real complex
systems of live and of lifeless things: in cardiology, physiology,
neurophysiology, biophysics of membranes and seismology.

\section{Acknowledgments} We gratefully acknowledge Prof. Raoul
Nigmatullin for stimulating discussion  and Dr.L.O.Svirina for
technical assistance. This work supported in part by Russian Fund
for Basic Research (Grant No 02-02-16146), Ministry of Education
of RF (Grant No A03-2.9-336, No 02-3.1-538) and RHSF (No
03-06-00218a).

%%%%%%%%%%%%%%%%%%%%%%%

\appendix \section{Three forms of the quasi-operator $\hat{\mathcal{L}}$} \label{A}

Equations of motion of a random variable $x$ with the use of the
derivative of the three different forms \cite{Korn} are
represented here.

Using evolution operator, we can write the equation of motion for
a discrete case as following \be
dx(t)/dt=i{\hat{\mathcal{L}}}x(t). \label{eq_of_m} \ee However,
there is a possibility of application of the time derivative $d/dt
\rightarrow \triangle / \triangle t$ in three different forms:

\textbf{1}. \textit{''Right'' derivative (with decurrent
difference in numerator)} \be \frac{\triangle x(t)}{\triangle
t}=\frac{x(t+\tau)-x(t)}{\tau}=\frac{1}{\tau}U(t+\tau,t)x(t)
\label{deriv_1} \ee with Liouville's quasioperator of the
following form: \be \hat{\mathcal{L}}(t,\tau)= -\frac{i}{\tau}
[U(t+\tau,t)-1]; \label{Liuv_1} \ee

\textbf{2}. \textit{''Left'' derivative (with ascending difference
in numerator)}
\be
\frac{\triangle x(t)}{\triangle t}=\frac{x(t)-x(t-\tau)}{\tau}=
\frac{x(t)-U^{-1}(t,t-\tau)x(t)}{\tau}=
\frac{1}{\tau}[1-U^{-1}(t,t-\tau)]x(t) \label{deriv_2}
\ee
with the Liouvillian of the next form:
\be
\hat{\mathcal{L}}(t,\tau)= -\frac{i}{\tau}[1-U^{-1}(t,t-\tau)];
\label{Liuv_2}
\ee

\textbf{3}. \textit{''Central'' derivative (with central
difference in numerator)}
\bn
\frac{\triangle x(t)}{\triangle
t}&=&\frac{x(t+\tau)-x(t-\tau)}{2\tau}=
\frac{x(t+\tau)-x(t)}{2\tau}-
\frac{x(t)-x(t-\tau)}{2\tau}\nonumber\\
&=&\frac{1}{2\tau}[U(t+\tau,t)-U^{-1}(t,t-\tau)]x(t).
\label{deriv_3}
\en
Then the quasioperator $\hat{\mathcal{L}}$ takes the following
form:
\be
\hat{\mathcal{L}}(t,\tau)=
-\frac{i}{2\tau}[U(t+\tau,t)-U^{-1}(t,t-\tau)]. \label{Liuv_3}
\ee

In the calculations and the analysis presented in this work we
have used the derivative of the first form [see Eqs.
\Ref{deriv_1}, \Ref{Liuv_1}].

\newpage
\begin{figure}
\centerline{\epsfig{figure=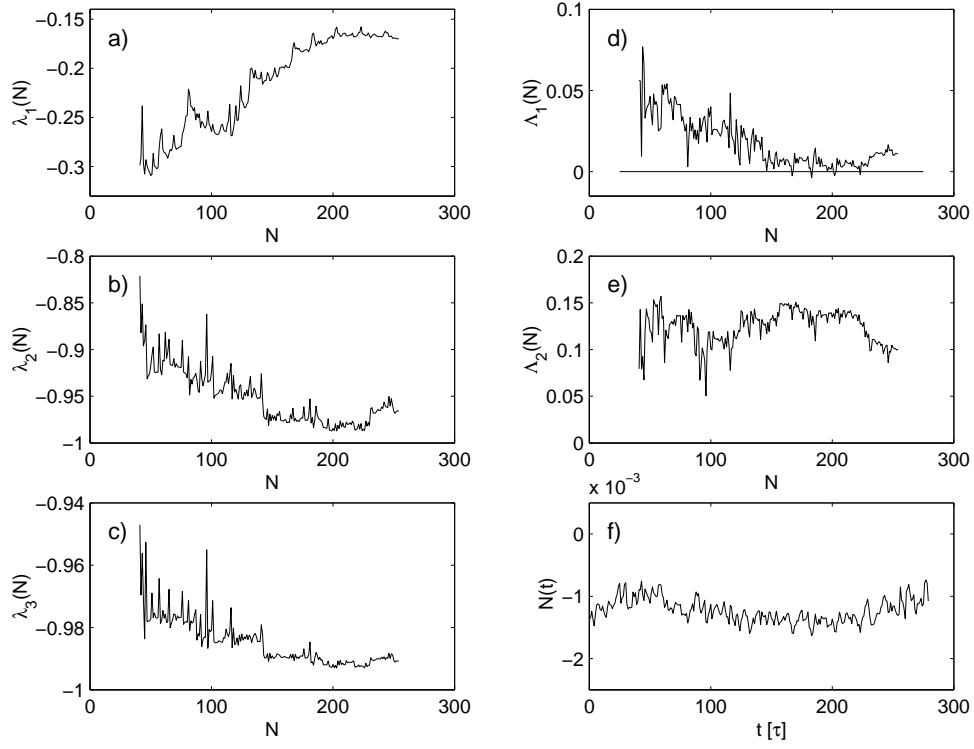,height=10cm,angle=0}}
\caption{The definition of the optimal size for a local "working
window". Calculations were made for the data of a steady state of
the Earth. As a result we find that size $N \sim 150$ points is
more optimal. The relative stability of all parameters is
observed. Inset f) shows seismic signals for the Earth's steady
state (before underground TE).}
\end{figure}

\begin{figure}
\centerline{\epsfig{figure=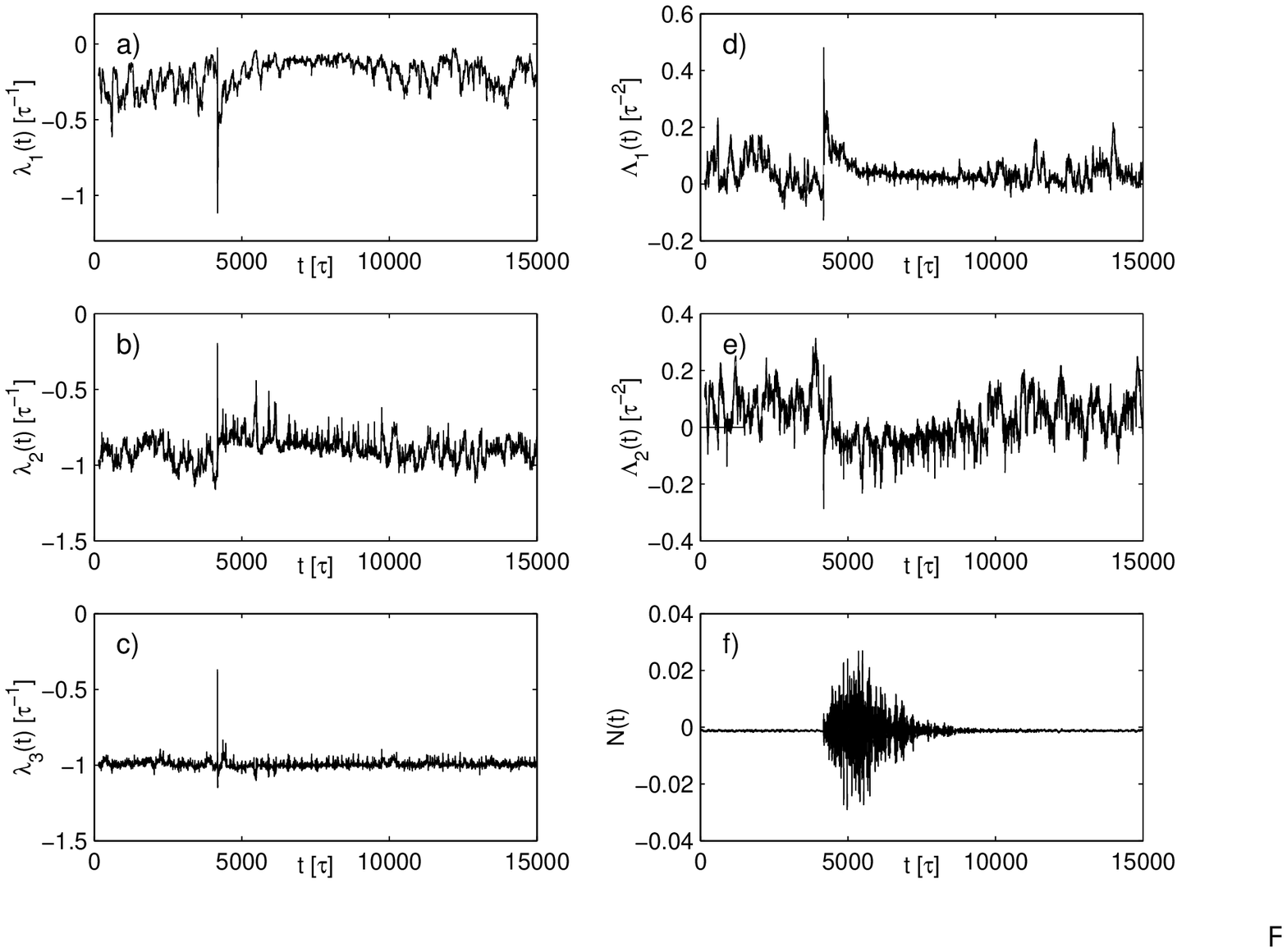,height=10cm,angle=0}}
\caption{The calculated time behavior of noisy relaxation
parameters $\lambda_{i}$ and $\Lambda_{i}$ for the technogenic
explosion TE(3) [figures a)-e)], the signal of which is presented
at inset f). }
\end{figure}

\begin{figure}
\centerline{\epsfig{figure=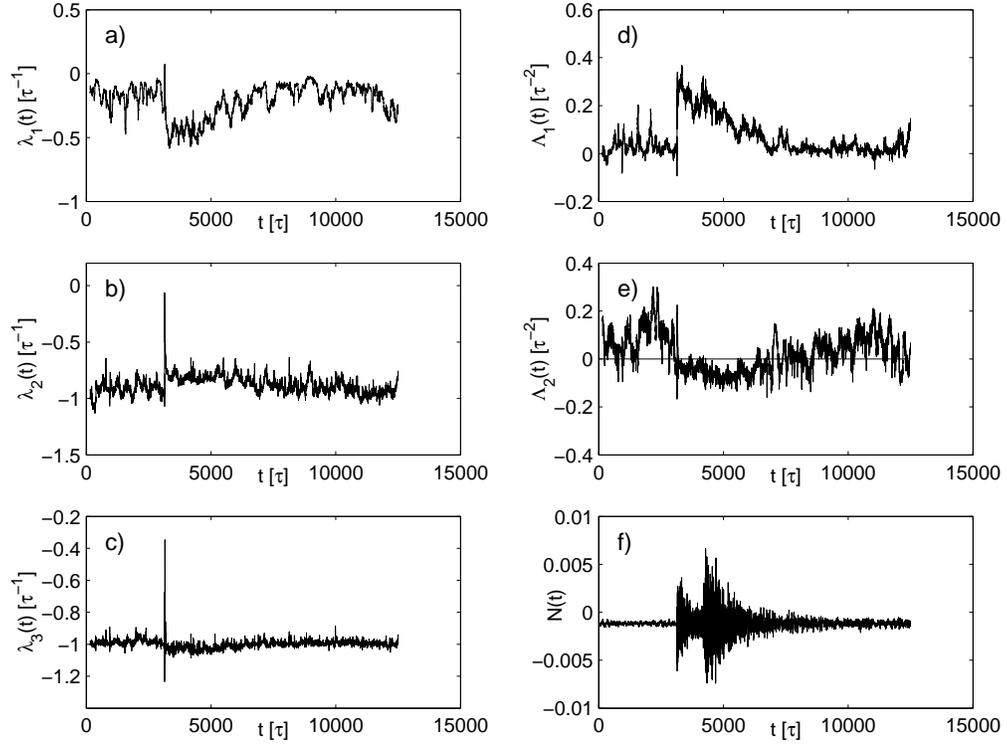,height=10cm,angle=0}}
\caption{The time behavior of the noise relaxation parameters
$\lambda_{i}$ and $\Lambda_{i}$ for the weak EQ(3) [see figures
a)-e)], the signal of which is also presented  at inset f).}
\end{figure}

\begin{figure}
\centerline{\epsfig{figure=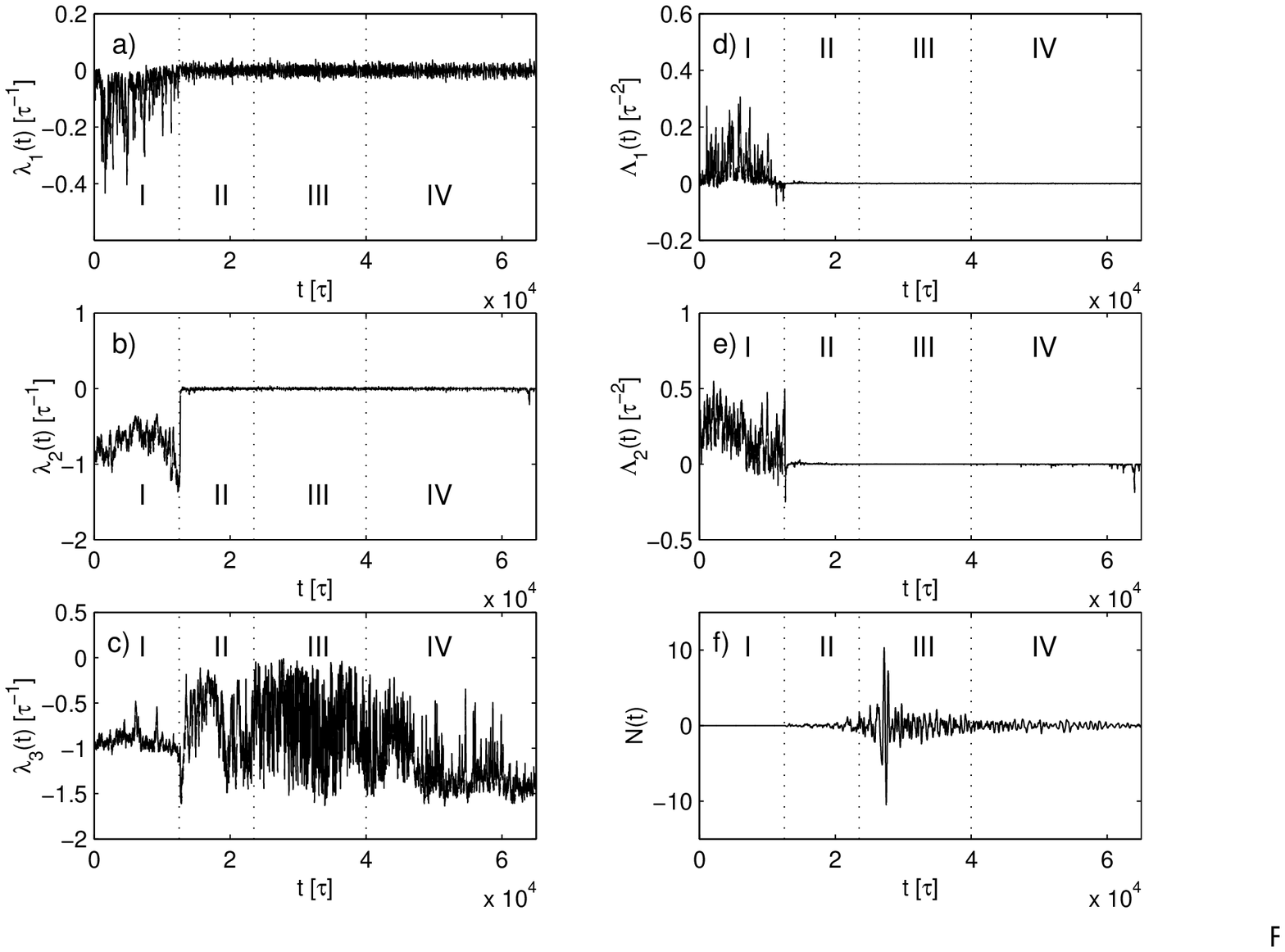,height=10cm,angle=0}}
\caption{The time behavior of the noisy relaxation parameters
$\lambda_{i}$ and $\Lambda_{i}$ [figures a)-e)] for the strong
EQ(T), the signal of which is presented in figure f). The
following regions: I - a calm state of the Earth's core, II - the
state before the earthquake, III - the state during the earthquake
and, finally, IV - the state after the event are divided by the
vertical dotted lines.}
\end{figure}

\begin{figure}
\centerline{\epsfig{figure=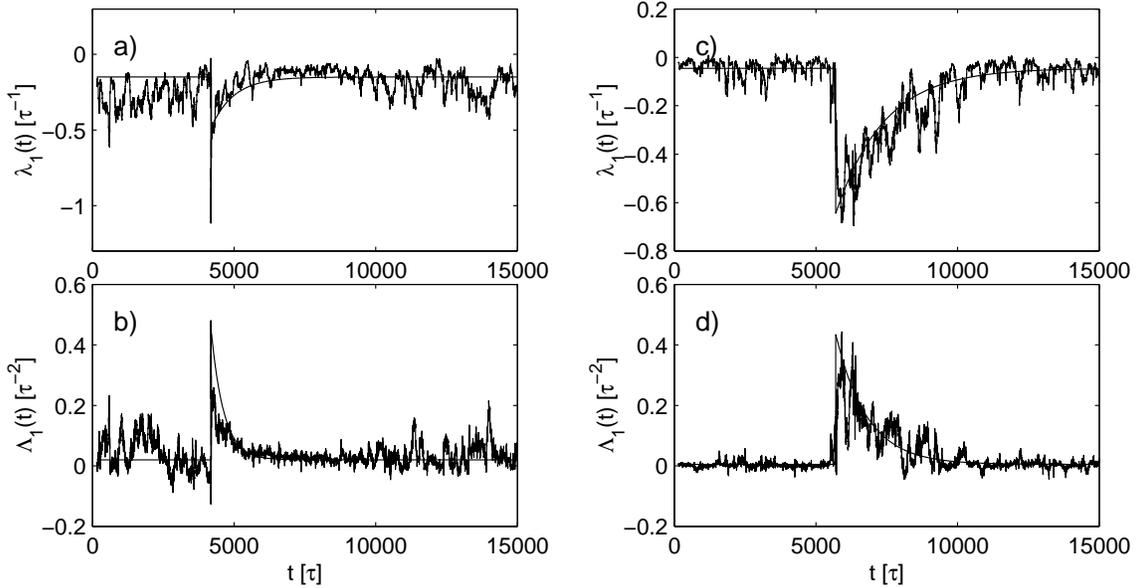,height=8cm,angle=0}}
\caption{The time behavior of the first two noisy relaxation
parameters $\lambda_{1}$, $\Lambda_{1}$ for TE(3) [figures a) and
b)] and for the weak EQ(1) [figures c) and d)]. Solid lines show
fitted functions \Ref{expon_1} and \Ref{expon_2} with
corresponding parameters, presented in Table II. One can note the
typical exponential restoration of the parameters with the
beginning of the event.}
\end{figure}


\begin{thebibliography}{99}
\bibitem{Herst} It should be noted the interest observation that local interval with
$100-120$ points was also found in previous work \cite{Yulmetyev1}
devoted to seismic data analysis with the help of local Herst
exponent.
\bibitem{param} It turned out, that the noise variation is visible
for all noise parameters $\lambda_{i}(t)$ and $\Lambda_{j}(t)$
($i=1,2,3;$ $j=1,2$) long before earthquake. However, it is more
pronounced for time-dependence of the parameter $\Lambda_{1}(t)$.
\bibitem{Turkish} The strong Turkish earthquake presented here was analized on
the basis of the data of different sources. However, we received
always results, which were the same with presented in Fig. 4.
\end{thebibliography}
\end{document}